\documentstyle[prl,aps,preprint,psfig]{revtex}

\begin{document}
\draft

\title{Magnetization of small lead particles}

\author{S. Reich$^1$, G. Leitus$^1$, R. Popovitz-Biro$^1$, 
and M. Schechter$^2$}

\address{$^1$Department of Materials and Interfaces, The 
Weizmann Institute of Science, Rehovot 76100, Israel 
\newline
$^2$ The Racah Institute of Physics, The Hebrew University, Jerusalem 
91904, Israel }

\maketitle

\begin{abstract}

The magnetization of an ensemble of isolated lead grains of sizes ranging 
from below $6$ nm to $1000$ nm is measured. A sharp disappearance of 
Meissner effect with lowering of the grain size is observed for the 
smaller grains. This is a direct observation by magnetization measurement 
of the occurrence of a critical particle size for superconductivity, 
which is consistent with Anderson's criterion.

\end{abstract}

\vspace{0.5cm}


In 1957 Bardeen, Cooper and Schrieffer (BCS) introduced their mechanism 
for superconductivity\cite{BCS57}, 
which gives an excellent description of low $T_{\rm c}$ 
superconductivity in bulk samples. Very soon afterwards, the question of 
the size dependence of superconductivity arose, and in $1959$ 
Anderson claimed\cite{And59} that for grains so small such that their 
level spacing $d$ 
is larger than the bulk gap $\Delta$, superconductivity would 
not exist, since such a grain will not have even one condensed level. 
Anderson's statement was particularly intriguing since it claims, on 
the other hand, that grains 
much smaller than the coherence length $\xi$ do have superconducting 
properties (the condition $d = \Delta$ is fulfilled at a grain of size 
$\xi^{(1/3)} \lambda_{\rm F}^{(2/3)} \ll \xi$ where $\lambda_{\rm F}$ is the 
Fermi wave length). Motivated by this statement, Giaver and Zeller 
\cite{GZ68,ZG69} measured the conductivity of small superconducting grains. 
They have indeed confirmed that grains much smaller than $\xi$ have a 
gap $\Delta$ in their single particle spectrum, but they could not confirm 
the loss of this property at smaller grains, in the regime where $d >\Delta$ 
(to be called the ultrasmall regime). Only recently Ralph Black and Tinkham 
(RBT)\cite{RBT97} have shown that a superconducting property persists when 
the size of the 
grain is reduced until the Anderson limit (where $d=\Delta$), but is lost 
in the ultrasmall regime. RBT have measured the tunneling spectrum of single 
Aluminum grains in the range of $3-5$ nm, and have shown that for the larger 
grains, for which $d < \Delta$, there exists a gap of $2 \Delta$ in the 
tunneling spectrum of grains with an odd number of particles. The smaller 
grains, with $d \approx \Delta$ did not show this property. This beautiful 
experiment initiated vast theoretical work, in which superconductivity of 
small grains was studied, and particularly the crossover between the regime 
where $d>\Delta$ and the ultrasmall regime (see \cite{DR01} and references 
therein). However, on the experimental 
side the work of RBT remained the single work which investigated the 
crossover across the Anderson limit. 

In this paper we present measurements of the magnetization of an
ensemble of $10^{12}$ Pb grains, ranging in size from $1000$ nm down
to grains smaller than $6$ nm. The larger grains, down to $30$
nm in size, showed diamagnetic response with the well known finite
size correction\cite{Sho52}.  However, at a smaller size, estimated to
be roughly $5$ nm, an abrupt change was found in the diamagnetic
response, which, within the experimental accuracy, vanishes for the
smaller grains. The above size is in agreement with the Anderson limit
for lead. Thus, our results suggest that grains with $d < \Delta$ show
the Meissner effect, and grains with $d > \Delta$ do not.  In the case
of RBT the loss of the superconducting property at the Anderson limit
was a direct consequence of the criterion itself, i.e. as soon as $d >
\Delta$ the superconducting gap can not be distinguished from the
gap due to the finite level spacing.  For the Meissner effect
measured here, the relation between the Anderson limit and the
existence of the superconducting property is less immediate, and
therefore of a deep physical meaning, reflecting the connection
between the superconducting correlations and the Meissner effect.  It
suggests that indeed, as long as there are even a few condensed levels,
their correlations suffice to create the Meissner effect, but as soon
as there is not even one condensed level, the Meissner effect
disappears.

A measurement of the magnetic response of small superconducting grains
required the possibility to produce a large ensemble of small grains,
with good size control, which are isolated one from the other. We
achieved this by using a method developed in our laboratory in 1990
\cite{KRM90}.  The lead particles are deposited into the pores of
polycarbonate nuclepore membranes (NP), and due to the confinement in the
pores the particles do not agglomerate. 
The deposition of lead particles into NP membranes was performed by
counter-current diffusion of a Pb-salt and a reducing agent from opposite
sides of the membranes (see experimental setup in Fig.\ref{figsetup}). 
These hydrophilic polyvinyl pyrrolidone (PVP) coated polycarbonate
membranes have straight -  through pores distributed randomly on the surface
and penetrating the surface at an angle of incidence $\pm 34^\circ$. 
The NP membranes
were equilibrated in triple distilled water for an hour; the membranes were
mounted into a cell shown in Fig.\ref{figsetup}.

Upon completion of the mounting
process, the two stirred chambers were filled with lead salt on the dull
side of the membrane and with reducing agent on the shining side. The
reducing agent was $0.2{\rm M}$ aqueous ${\rm NaBH_4}$ solution. The lead salt 
solution was $80{\rm cc}$ of $0.03{\rm M \; PbNO_3} + 20{\rm cc}$ of 
$1{\rm M \; HNO_3}$. These two solutions were introduced
simultaneously to the two chambers of the diffusion cell. The deposition
time varied with the pore size of the membranes and the desired lead loading
into the pores. This time varied from a few minutes for a full loading of a
$1000$ nm membrane to a few hours for a $10$ nm NP.
Upon completion of the deposition process, both chambers were emptied and
filled with absolute alcohol for one minute. The NP membranes were then
dried on blotting paper. Note that the lead salt was introduced in an acidic 
solution to prevent the formation of hydroxide species which may lead to the 
formation of lead oxide. To reveal the morphology of the lead particles 
in the pores of the NP membranes 
TEM (transmission electron microscope) imaging was carried out. 
NP membranes were embedded in epoxy resin 
and sectioned into thin ($50-70$nm) slices. 
Slices were cut parallel
to the NP membrane surface, thus approximately perpendicular to the pores
in the membranes. Fig. \ref{figparticles} shows such slices through the 
membranes for few NP
diameters. In Fig. \ref{figslice} we show a slice which is almost parallel 
to the pores of the NP.
The TEM micrographs show good control of the diameter of the embedded
particles. However the particle length, especially at high loading, varies
as shown in Fig. \ref{figslice}.
The lead weight content per unit surface of a membrane was determined by
inductively coupled plasma emission spectroscopy (ICP) after digestion of
lead into a $65 \% \; {\rm HNO_3}$ solution.
XRD characterization of the lead loaded NP membranes showed the presence of
metallic lead.

Magnetization measurements were performed with a ${\rm MPMS_2}$ field screened
magnetometer. All magnetization curves presented are corrected for the
diamagnetic contribution of the polymeric membranes. For the larger 
grains, of sizes $30-1000$ nm, we obtain the known magnetization curves 
of small superconducting particles\cite{Sho52,Pon37}. 
In Fig. \ref{figpon} we present Magnetization ($M$) vs. magnetic field 
($H$) curves at $5K$ for 
few pore sizes in the above range. 
These measurements were performed with $H$ 
parallel to the membrane surface. The data were normalized according to 
Eq.~(\ref{norm}), assuming that the difference between the free 
energy density $g_n - g_s$ is independent of the size of the particles;

\begin{equation}
g_n - g_s = - \int_0^h M dH_e = A
\label{norm}
\end{equation}
where $H_e$ is the external field and $A$ is the area under the 
magnetization curve \cite{Pon37}. 
Since the slope of the magnetization curve is smaller for a small
specimen than for a bulk one of the same shape, the magnetization 
continues to higher fields to have the same area, i.e. the critical
field, $h$, as determined from the intercept of the tangent to
the rising branch of magnetization and the horizontal field independent
branch, shifts to higher fields for smaller particles. We observe this 
shift, as well as a shift 
of the magnetization minimum to
higher field values upon the decrease in the diameter of the particles. 
Plotting $h/H_c$ as function of $1/R$, where $R$ is the radius of the 
pores in the NP membrane and $H_c$ is the critical field for the bulk 
lead ($H_c = 415$ Oe at $5K$), we obtain (see insert of Fig~\ref{figpon}) 
the relationship\cite{Sho52,Pon37} 

\begin{equation}
\frac{h}{H_c} = 1 + \frac{b}{R}
\label{eqpon}
\end{equation}
with $b=7.62 \cdot 10^{-6}$ cm 
(for the data in Ref.~\cite{Pon37} a similar relation 
was found, with $b=11 \cdot 10^{-6}$ cm\cite{Sho52}).

However, for smaller grains, of sizes $10$ nm and below, we find a very 
different behavior of $M$ vs. $H$. We use a membrane with $10$ nm pores, 
and control the size of the lead grains inside these pores by varying the 
deposition time. The content of lead per unit area of the membranes was 
determined by the ICP technique. 
We find that below $5 \, {\rm \mu g/cm}^2$ loading no Meissner 
effect is observed, see Fig.~\ref{figsmall}. 
The transition, as function of size, between the 
regime where Meissner effect is observed and the regime where it is not 
observed is sharp, as can be seen in the insert of Fig.~\ref{figsmall}, 
where the integral values under the magnetization curves as function of the 
lead loadings are drawn. We estimate the size of the grains for which 
the transition occurs to be below $6$ nm. This estimate is consistent 
with the condition $d = \Delta$ for lead particles, which gives an 
approximate size of $5$ nm. This is the central result of this paper. 

Another observation for these grain sizes is that the critical field, $h$, 
does not obey Eq.~(\ref{eqpon}), but is considerably smaller. We do not 
have a clear understanding of the physics responsible for this observation. 
We would like to mention, in this regard, that recent 
experimental\cite{HWB+01} and 
numerical\cite{HH01} works have found that at grain sizes of this 
order, lead grains change their crystalline structure due to the large 
portion of surface atoms. Such a change could affect the superconducting 
properties of the grain. We would like to stress that the lowering of 
the critical field and the loss of the Meissner effect occur at different 
grain sizes, and are therefore clearly two different phenomena. 
 

\begin{thebibliography}{10}

\bibitem{BCS57}
J. Bardeen, L.~N. Cooper, and J.~R. Schrieffer, Phys. Rev. {\bf 108},  1175
  (1957).

\bibitem{And59}
P.~W. Anderson, J. Phys. Chem. Solids {\bf 11},  26  (1959).

\bibitem{GZ68}
I. Giaever and H.~R. Zeller, Phys. Rev. Lett. {\bf 20},  1504  (1968).

\bibitem{ZG69}
H.~R. Zeller and I. Giaever, Phys. Rev. {\bf 181},  789  (1969).

\bibitem{RBT97}
D.~C. Ralph, C.~T. Black, and M. Tinkham, Phys. Rev. Lett. {\bf 78},  4087
  (1997).

\bibitem{DR01}
J. von Delft and D. Ralph, Phys. Rep. {\bf 345},  61  (2001).

\bibitem{Sho52}
D. Shoenberg, {\em Superconductivity} (Cambridge University press, Cambridge,
  1952).

\bibitem{KRM90}
D. Katzen, S. Reich, and S. Mazur, Jour. of Materials Science {\bf 25},  2056
  (1990).

\bibitem{Pon37}
R.~B. Pontius, Phil. Mag. {\bf 24},  787  (1937).

\bibitem{HWB+01}
M. Hyslop, A. Wurl, S.~A. Brown, B.~D. Hall, and R. Monot, Eur. Phys. J. D {\bf
  16},  233  (2001).

\bibitem{HH01}
S.~C. Hendy and B.~D. Hall, Phys. Rev. B {\bf 64},  085425  (2001).

\end{thebibliography}

\begin{figure} 
	\caption{Diffusion Cell. 1: PMMA transparent body. 2 and 4: pressure 
	clamp. 3: magnetic stirrer. 5 and 6: shaft assembly for the 
	stirrers. 7: NP membrane.  }
	\label{figsetup}
\end{figure}

\begin{figure} 
	\caption{TEM micrographs, slices parallel to the membrane for 
	(A) $10$ nm membrane loaded with lead (B) $30$ nm NP membrane 
	(C) $50$ nm NP membrane (D) $100$ nm NP membrane.  }
	\label{figparticles}
\end{figure}

\begin{figure} 
	\caption{TEM micrograph for 50 nm NP membrane, slice perpendicular 
	to the membrane. } 
	\label{figslice}
\end{figure}

\begin{figure} 
	\caption{Normalized magnetization vs. $H$ at $5K$ according to 
	Eq.~\ref{norm} for different pore size NP membranes loaded with 
	lead. Insert: normalized critical fields, $h/H_c$ vs. $1/R$ (see 
	text).  } 
	\label{figpon}
\end{figure}

\begin{figure} 
	\caption{$M$ vs. $H$ at $5K$ for different loads 
	of lead into $10$ nm NP membranes. Insert: the integral under the 
 	magnetization curves in the field limits $0-4000$ Oe, as function 
	of lead loading into pores. Graphs correspond to lead loads and 
	deposition times as follows: (i) $1.2 \, {\rm \mu g/cm}^2, 0.5$ hours 
	(ii)  $3.9 \, {\rm \mu g/cm}^2, 1.0$ hours 
	(iii)  $9.1 \, {\rm \mu g/cm}^2, 2.0$ hours 
	(iv) $11.7 \, {\rm \mu g/cm}^2, 2.5$ hours 
	(v) $14.6 \, {\rm \mu g/cm}^2, 3.0$ hours 
	(vi) $19.6 \, {\rm \mu g/cm}^2, 4.0$ hours  } 
	\label{figsmall}
\end{figure}


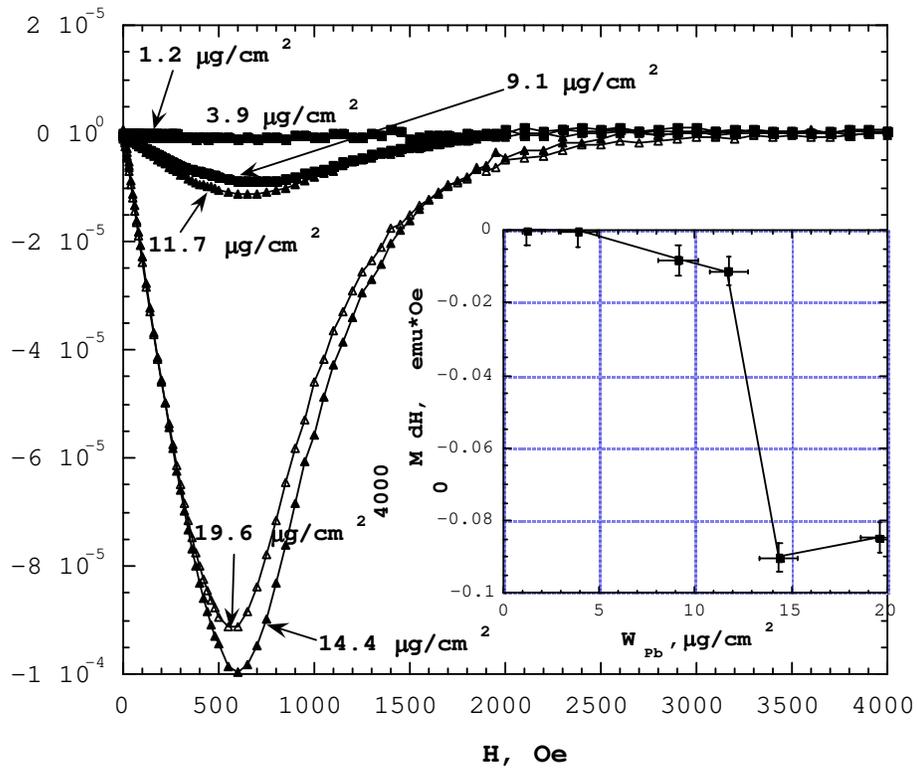
\end{document}